\documentclass[conference]{IEEEtran}
\IEEEoverridecommandlockouts
\usepackage{amsmath,amsfonts}
\usepackage{algorithmic}
\usepackage{textcomp}
\usepackage{stfloats}
\usepackage{algorithm}
\usepackage{array}
\usepackage[caption=false,font=normalsize,labelfont=sf,textfont=sf]{subfig}
\usepackage{textcomp}
\usepackage{stfloats}
\usepackage{url}
\usepackage{verbatim}
\usepackage{graphicx}
\usepackage{caption}
\usepackage{romannum}
\usepackage{cite}
\usepackage{amsmath}
\usepackage{makecell}
\usepackage{multirow}
\usepackage{xcolor}
\usepackage{gensymb}
\newcommand{\norm}[1]{\left\lVert #1 \right\rVert}

\begin{document}

\title{A Virtual Admittance-Based Fault Current Limiting Method for Grid-Forming Inverters

\thanks{This research was supported by the National Science Foundation through award number 1946093.}
}

\vspace{-8 pt}
\author{
\IEEEauthorblockN{Zaid Ibn Mahmood}
\IEEEauthorblockA{\textit{School of ECE} \\
Oklahoma State University, U.S. \\
zaid.mahmood@okstate.edu}
\and
\IEEEauthorblockN{Hantao Cui}
\IEEEauthorblockA{
\textit{Department of ECE} \\
North Carolina State University, U.S. \\
hcui9@ncsu.edu}
\and
\IEEEauthorblockN{Ying Zhang}
\IEEEauthorblockA{
\textit{School of ECE} \\
Oklahoma State University, U.S. \\
y.zhang@okstate.edu}
}

% \author{Zaid Ibn Mahmood,~\IEEEmembership{Graduate Student Member,~IEEE,}
%         Hantao Cui,~\IEEEmembership{Senior Member,~IEEE},\\
%         Ying Zhang,~\IEEEmembership{Member,~IEEE}
%         % <-this % stops a space
% \vspace{-25pt}
% \thanks{Z. Mahmood and Y. Zhang are with the School of Electrical and Computer Engineering, Oklahoma State University, Stillwater, OK 74074, USA (e-mail:zaid.mahmood@okstate.edu and  y.zhang@okstate.edu). 

% H. Cui is with the Department of Electrical and Computer Engineering, North Carolina State University,  Raleigh, NC 27606, USA (e-mail: hcui9@ncsu.edu)

% This research was supported by the National Science Foundation through award number 1946093.}}% <-this % stops a space

\maketitle

\begin{abstract}
Inverter-based resources (IBRs) are a key component in the ongoing modernization of power systems, with grid-forming (GFM) inverters playing a central role. Effective fault current limiting is a major challenge to modernizing power systems through increased penetration of GFM inverters. Due to their voltage-source nature, GFM inverters offer no direct control over the output current and, therefore, are susceptible to high fault currents. This vulnerability is especially pronounced during large phase jumps, a condition overlooked by most fault current limiting methods. This paper proposes a hybrid fault current limiting method implemented through a virtual admittance by leveraging the advantages of two virtual impedance (VI)-based methods tailored for three-phase faults and phase jump disturbances. Electromagnetic transient simulations conducted in MATLAB-Simulink demonstrate the method’s effectiveness across various disturbances, validating its potential in single-loop GFM structures.

\end{abstract}

\vspace{-10pt}
\section{Introduction}

\IEEEPARstart{T}{he} growing demand for cleaner energy necessitates modernizing the power system by integrating more renewable energy sources \cite{milanoFoundationsChallengesLowInertia2018}. DC/AC inverters are key components in this transition as they interface renewable energy sources to the power grid. With the ability to establish a voltage at the grid terminal, grid-forming (GFM) inverters are particularly essential for power systems with high renewable penetration. Due to their voltage-source nature, GFM inverters concede direct control over the output phase current. Therefore, a large magnitude of currents can flow through them, damaging these power-electronic devices. This paper addresses this challenge by proposing a fault current limiting method, which is robust across various disturbance scenarios.  

Synchronous generators can withstand an overcurrent of 5–7 per unit ($p.u.$), whereas semiconductor-based inverters typically tolerate only 1.2–2 $p.u.$ for a much shorter period  \cite{fanReviewCurrentLimitingControl2022}. Compared with synchronous generators, the semiconductor-based units in GFM inverters inherently limit their ability to tolerate overcurrent conditions.  This limitation underscores the need for robust fault current limiting methods for GFM inverters in high-penetration renewable systems.

% \subsection{Literature Review}
The existing current-limiting methods can be broadly classified into two categories: current limiters and virtual impedance-based limiters \cite{zengHybridThresholdVirtual2023}.
Current limiters restrict current references whenever they exceed a preset threshold value, effectively limiting the output current. However, due to the use of limiters, current limiter methods can lead to integrator windup in the proportional-integral (PI) controllers and prevent successful fault recovery \cite{fanReviewCurrentLimitingControl2022}. Additionally, they can distort voltage signals as currents are clipped, leading to reduced power quality \cite{zareiReinforcingFaultRide2019}.  
On the other hand, virtual impedance-based methods adjust the output voltage reference by introducing a voltage drop via a virtual impedance. In the standard threshold virtual impedance (TVI) method, the virtual voltage drop is proportional to the phase current magnitude \cite{paquetteVirtualImpedanceCurrent2015}. Several improvements have been suggested to improve the efficiency of the TVI method. To improve the speed of TVI by achieving asymptotic current tracking, a dynamic threshold-based method is proposed in \cite{shiraziEvaluationCurrentLimitingStrategies2023}. Further, a TVI method is proposed in \cite{wangCurrentLimitingSchemeVoltageControlled2023}, which utilizes instantaneous current measurements, reducing activation time by skipping current magnitude calculation. In \cite{qoriaVariableVirtualImpedanceBased2023}, a variable TVI method is proposed, which implements a transient resistance to balance current oscillation damping and sufficient transient stability. Yet, these methods are designed for three-phase fault scenarios, and they might not be equally effective for other disturbances, such as phase jumps where voltage information is crucial.

Phase jumps or phase shifts are sudden shifts in voltage angle that can result from events such as three-phase faults, heavy load connections, or line tripping and reclosing \cite{qoriaVariableVirtualImpedanceBased2023}. Depending on the magnitude of the phase jump propagated to the inverter terminal, traditional TVI methods based only on the phase current magnitude, such as \cite{paquetteVirtualImpedanceCurrent2015}, might not be effective as they do not consider cases where the voltage drop across the virtual impedance exceeds nominal inverter voltage. To solve this, the work in \cite{zengHybridThresholdVirtual2023} proposes a voltage information-based virtual impedance (VIv) method taking into consideration terminal voltage information, which is suitable for the dual-loop GFM structure. However, it has been observed that it is not equally suited for single-loop GFM structures due to its reliance on measuring current derivatives. 

In summary, the majority of virtual impedance-based current limiting methods are not equipped to limit current effectively in phase shift scenarios. Although the method in \cite{zengHybridThresholdVirtual2023} is designed to handle phase jump or shift scenarios, it cannot be implemented in a single-loop GFM structure due to difficulty in achieving current derivatives.

% \subsection{Contribution}

To address the highlighted issues, this paper proposes a phase information-aware virtual admittance-based fault current limiting method, which is robust for both three-phase faults and phase jump scenarios. A virtual impedance requires differentiating current measurements, which often leads to instability. On the contrary, a virtual admittance-based approach is essentially implemented through a low-pass filter, avoiding the issues with differentiating. The proposed method adapts the approach in \cite{zengHybridThresholdVirtual2023} to a virtual admittance framework, making it compatible with single-loop GFM structures.

The rest of the paper is organized as follows: Section II introduces the technical background. Then, section III introduces the proposed current limiting method. Further, section IV provides simulation results to highlight the effectiveness of the proposed method in both three-phase fault and phase jump disturbances. Finally, conclusions are drawn in Section V.
\vspace{-10 pt}
\section{System Overview and Technical Background}
\subsection{Overview of the GFM Control Diagram}

\begin{figure}
    \centering
    \includegraphics[width=\linewidth]{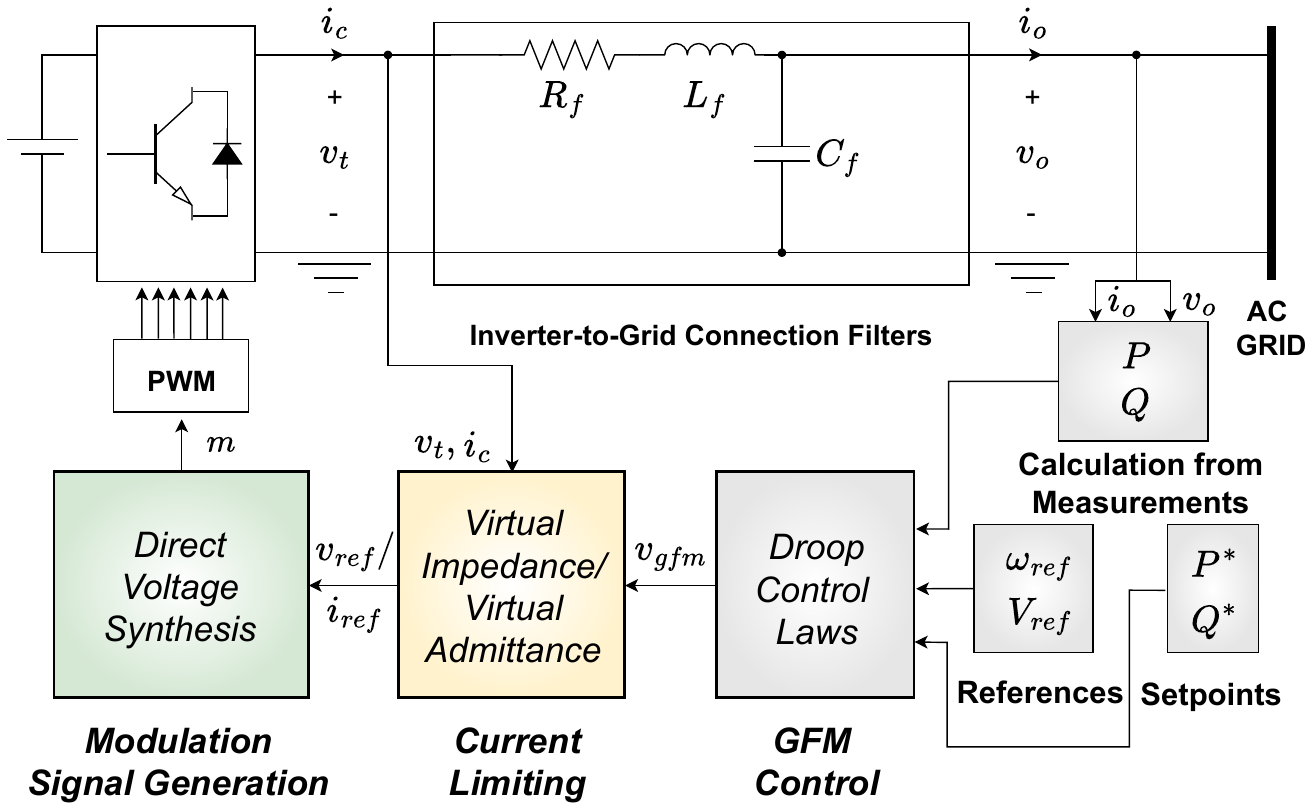}
    \caption{Control diagram of a GFM inverter with current limiting capabilities connected to the AC Grid.}
    \label{fig:control}
\end{figure}

Fig. \ref{fig:control} illustrates the system layout of a GFM inverter connected to an AC grid. The inverter is linked to the grid via an RLC filter containing a filter resistor with resistance $R_f$, a filter inductor with inductance $L_f$, and a filter capacitor with capacitance $C_f$. As common in current limiting studies, the DC-link voltage is considered constant. GFM phase current and terminal voltage are represented by $i_c$ and $v_t$, while the filter output current and voltage are represented by $i_o$ and $v_o$.

This paper implements the most widely implemented droop control law for GFM reference control. The inverter takes references $V_{ref}$ and $\omega_{ref}$ as voltage magnitude and frequency references, while $P^*$ and $Q^*$ are active and reactive power setpoints. The output active and reactive powers, $P$ and $Q$, are controlled by adjusting the generated references, $V_{gfm}$ and $\omega_{gfm}$, via droop laws as given in \eqref{droop}. Low-pass filters (LPF) with cut-off frequency $\omega_f$ are used to calculate the average active and reactive power from grid measurement values. 
\begin{subequations}
\label{droop}
\begin{align}
    \omega_{gfm} = \omega_{ref} + m_{p}(P^*-P \frac{\omega_f}{\omega_f + s}) \label{eq1a} \\
    V_{gfm} = V_{ref} + m_{q}(Q^*-Q\frac{\omega_f}{\omega_f + s}) \label{eq1b}
\end{align}
\end{subequations}
where $m_p$ and $m_q$ are the $P-\omega$ and $Q-V$ droop coefficients.

Current limiting can be implemented in two ways. In virtual impedance-based methods, $v_{gfm}$ is adjusted to $v_{ref}$ based on the virtual voltage drop. On the other hand, in virtual admittance-based methods, the current reference, $i_{ref}$, is generated based on $v_{gfm}$. 

Finally, a modulation signal, $m$, is generated to establish the terminal voltage, $v_t$, based on the references. This paper implements the single-loop direct voltage synthesis method to generate $m$.

\subsection{Review of Virtual Impedance-based 
Current Limiting Methods}

\begin{figure}[!t]
    \centering
    \includegraphics[width=.8\linewidth]{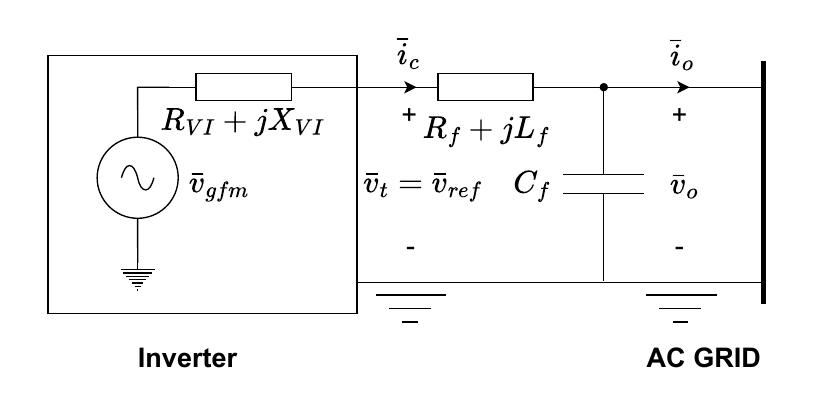}
    \caption{Simplified quasi-steady-state model of a GFM converter with virtual impedance-based current limiting.}
    \label{fig:enter-label}
        \vspace{-10 pt}
\end{figure}

This section briefly reviews two virtual impedance-based current limiting methods and a hybrid method that leverages their respective advantages. It then highlights the limitations of this approach when applied to the single-loop GFM structure.

\subsubsection{Threshold Virtual Impedance}

Fig. 2 shows a simplified quasi-steady-state model of a GFM inverter with virtual impedance, with filter dynamics neglected. In TVI, the resistive component, $R_{TVI}$, and reactive component, $X_{TVI}$, are designed to limit the phase current magnitude, $\norm{i_c}$, to maximum allowable current, $I_{max}$.
\begin{subequations}
\label{TVI}
\begin{align}
    R_{TVI} &:= 
    \begin{cases}
    0 & \norm{i_c} < I_{th} \\
    k_R(\norm{i_c} - I_{th}) & \norm{i_c} \geq I_{th}
    \end{cases}
    \\ 
    X_{TVI} &:= \sigma R_{TVI}
\end{align}
\end{subequations}
where $\sigma$ is the $X/R$ ratio, and $I_{th}$ is the threshold value of the phase current that activates the virtual impedance. 
\begin{equation}
    \label{kr}
    k_R := \frac{V_n}{I_{max}(I_{max}- I_{th}) \sqrt{\sigma^2 + 1}}
\end{equation}
where $V_n$ is the nominal inverter voltage. 

With TVI active, and neglecting the filters \cite{fanReviewCurrentLimitingControl2022}, the quasi-steady-state phase current magnitude, $\norm{\Bar{i_c}}$, can be expressed by \eqref{TVI_i}.
\begin{equation}
    \label{TVI_i}
    \norm{\Bar{i_c}} = \frac{\norm{\Bar{{v}}_{gfm} - \Bar{{v}}_{o}}}{\sqrt{R_{TVI}^2 + X_{TVI}^2}}
\end{equation}

The TVI impedance, $Z_{TVI} = \sqrt{R_{TVI}^2 + X_{TVI}^2}$, is limited by the maximum value of $\norm{\Bar{{v}}_{gfm} - \Bar{{v}}_{o}}$. Without sufficient phase difference between $\Bar{v}_{gfm}$ and $\Bar{v}_o$, this maximum value is equal to $V_n$, when the fault is at the inverter terminal, causing $v_o = 0$. In this extreme scenario, $\norm{\Bar{i_c}}$ is limited to $I_{max}$, ensuring reliable fault current limiting. Since, in most three-phase faults, the phase shift propagating to the inverter terminal is not too large, the TVI method remains reliable even under the most severe three-phase fault conditions. Yet, if the phase shift is large enough so that $\norm{\Bar{{v}}_{gfm} - \Bar{{v}}_{o}} > V_n$, the TVI method fails to restrain the phase current to safe levels. 

\subsubsection{Virtual Impedance Based on Voltage Information}

A voltage phase information-aware virtual impedance method, known as VIv, is introduced for reliable current limitation during phase jump scenarios. This method introduces virtual voltage drops based on voltage information under overcurrent conditions. The resistive and reactive components are generated according to \eqref{VIv}.
\begin{subequations}
\label{VIv}
\begin{align}
    R_{VIv} &:= 
    \begin{cases}
    0 & \norm{i_c} < I_{th} \\
    \frac{1}{\sqrt{\sigma^2 + 1}}(\frac{\norm{v_{gfm} - v_{o}}}{I_{max}}) & \norm{i_c} \geq I_{th}
    \end{cases}
    \\ 
    X_{VIv} &:= \sigma R_{VIv}
\end{align}
\end{subequations}

Under overcurrent conditions, $\norm{i_c}$ is limited to $I_{max}$, regardless of the phase difference between $v_{gfm}$ and $v_o$ by.
\begin{equation}
    \label{VIv_i}
    \norm{\Bar{i_c}} = \frac{\norm{\Bar{{v}}_{gfm} - \Bar{{v}}_{o}}}{\sqrt{R_{VIv}^2 + X_{VIv}^2}} = I_{max}
\end{equation}

\subsubsection{Hybrid Threshold Virtual Impedance}

Both the TVI and VIv methods have distinct advantages and drawbacks. The TVI method struggles to limit phase current within acceptable levels during large phase jumps. On the other hand, it is faster than the VIv method in current limiting and fault clearing during three-phase faults.

To combine the strengths of TVI and VIv methods, a hybrid threshold virtual impedance (HTVI) approach is proposed in \cite{zengHybridThresholdVirtual2023}. This method continuously evaluates the virtual impedances of both TVI and VIv during overcurrent conditions and selects a virtual impedance approach based on the maximum impedance. It is proposed in \cite{zengHybridThresholdVirtual2023} that $Z_{TVI} > Z_{VIv}$ when $\norm{\Bar{{v}}_{gfm} - \Bar{{v}}_{o}} > V_n$, and vice-versa when $\norm{\Bar{{v}}_{gfm} - \Bar{{v}}_{o}} < V_n$. Thus, the virtual impedance is chosen as the criterion for selecting the appropriate method to ensure reliable current limiting across various disturbances.

\subsection{Limitations of Virtual Impedance-based Methods}

The HTVI method is suitable for application with the cascaded dual-loop GFM structure as the structure inherently assumes a quasi-steady-state approximation. The virtual voltage drops are calculated in $d-q$ domains and subtracted from the reference voltage $d$ and $q$ components. However, the single-loop GFM structure does not assume a quasi-steady-state approximation. 

In the virtual impedance methods implemented with the single-loop GFM structure, the virtual voltage drop, $v_{vi}$, is calculated in the Laplace domain as shown in \eqref{VI}.
\begin{equation}
    \label{VI}
    v_{vi}(s) = v_{gfm}(s) - v_{t}(s) = i_c (s) (R_{VI} + s L_{VI})
\end{equation}

As observed, the inductive part of $v_{vi}(s)$ requires differentiation of $i_c$, which is challenging in practical cases due to noise. This is more prominent in fault scenarios as the current has a very large slew rate at the fault inception, potentially leading to instability. 

Two solutions are proposed in the literature to obtain the inductive voltage drop \cite{rodriguezControlPVGeneration2013}. The first uses LPFs to smooth the current waveform before differentiation. Yet, this introduces phase shifts and delays. Further, the choice of LPF cutoff frequency is not straightforward and often heuristic \cite{wangVirtualImpedanceBasedControlVoltageSource2015}. Higher cutoff frequencies reduce delay but increase harmonic content, while lower frequencies may lead to instability. The second solution avoids differentiation through the quasi-stationary approximation. Yet, this approximation only guarantees proper operation at the fundamental frequency.

\section{Proposed Virtual Admittance-based Current Limiting Method}

\begin{figure}[!t]
    \centering
    \includegraphics[width=\linewidth]{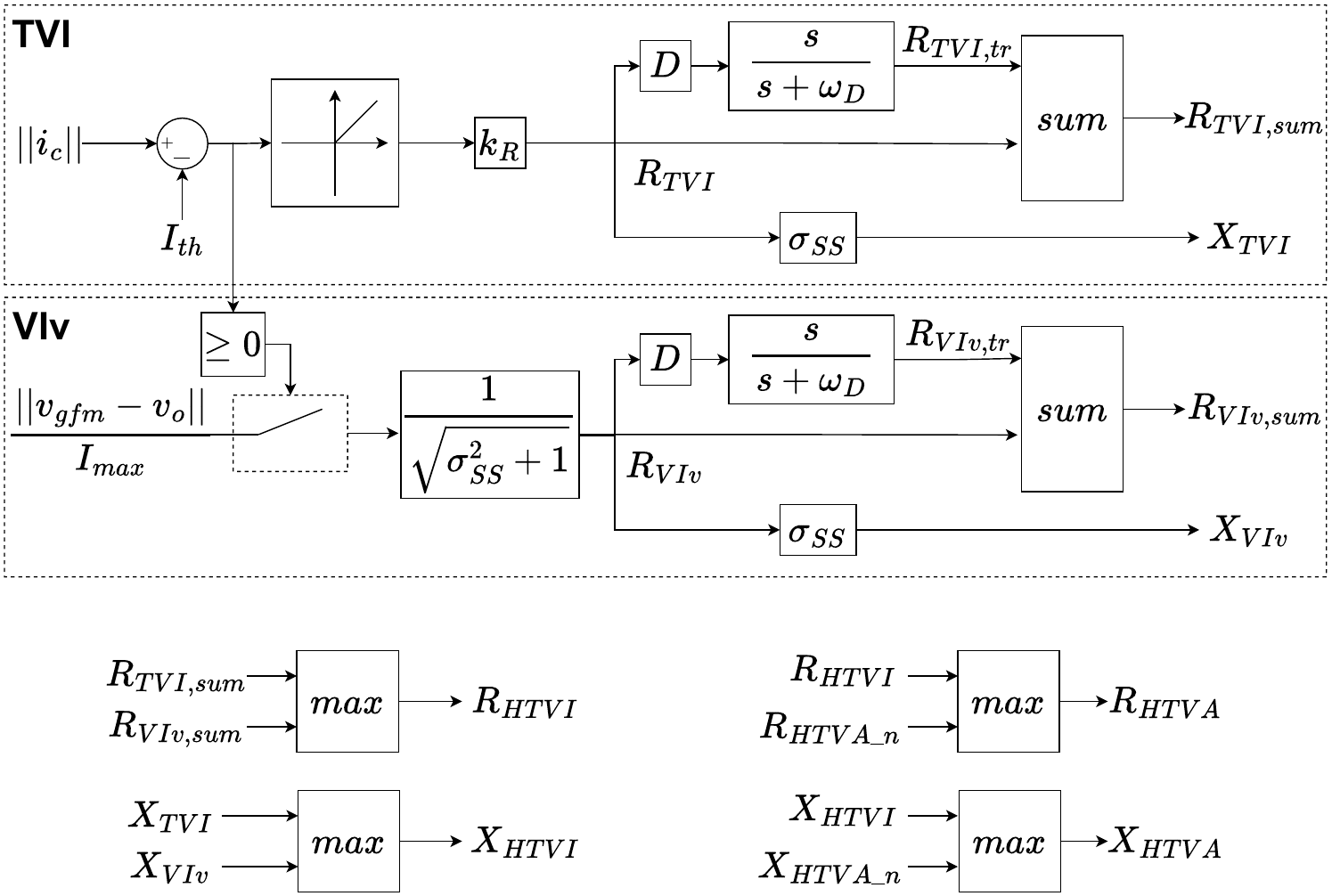}
    \caption{Control diagram of the proposed HTVA-based current limiting.}
    \label{fig:HTVA}
\end{figure}

The virtual admittance approach is implemented from a re-organization of \eqref{VI} as shown in \eqref{VA}.
\begin{equation}
    \label{VA}
     i_c (s) = \frac{1}{R_{VI} + s L_{VI}}(v_{gfm}(s) - v_{t}(s))
\end{equation}

The differentiation of current measurement can be avoided by implementing a virtual admittance-based on \eqref{VA}. Current limiting can be ensured by proper adjustment of the admittance $\frac{1}{R_{VI} + s L_{VI}}$.

The proposed hybrid threshold virtual admittance (HTVA) method incorporates resistive and inductive elements derived from the HTVI method, specifically adapted to a virtual admittance structure. This design ensures reliable current limiting for both three-phase faults and phase jump scenarios, while also overcoming the limitations of the virtual impedance method. A current reference, $i_{ref}$, is generated and fed to a current controller to ensure that $i_c$ tracks $i_{ref}$. 
\begin{equation}
    \label{VA_iref}
     i_{ref} (s) = \frac{v_{gfm}(s) - v_{t}(s)}{R_{HTVA} + s L_{HTVA}}
\end{equation}
During disturbances, virtual resistance, $R_{HTVA}$, and inductance, $L_{HTVA}$, are simply equal to their HTVI counterparts, which are determined as shown in Fig. \ref{fig:HTVA}. This paper employs the variable transient virtual resistance (VTRV) proposed in \cite{qoriaVariableVirtualImpedanceBased2023}, along with the HTVI. A large $\sigma \geq 5$ ensures better transient stability but may lead to an oscillatory current, while a low $\sigma$ yields a well-damped response but limited stability. To balance these effects, a dynamically adjusted $\sigma$ is proposed, where a high-pass filter (HPF) with cut-off frequency $\omega_D$ and damping factor $D$ is employed to activate resistance $R_{VI, tr}$ to reduce $\sigma$ during transients.
\begin{align}
    \sigma(s) &= \frac{X_{VI}}{R_{VI} + R_{VI,tr}} & D &= \frac{\sigma_{SS}}{\sigma_{TR}} - 1
\end{align}
where $\sigma_{SS}$ and $\sigma_{SS}$ denote the steady-state and transient X/R ratio. A detailed formulation can be found in \cite{qoriaVariableVirtualImpedanceBased2023}.

On the other hand, during steady-state, the impedance cannot be set to zero in the virtual admittance structure. Thus, $R_{HTVA}$ and $L_{HTVA}$ are set to nominal values $R_{HTVA\_n} = 0.1\ p.u.$ and $L_{HTVA\_n} = 0.2\ p.u.$.

\section{Test Cases}

The current limiting capability of the proposed HTVA method is evaluated through electromagnetic transient simulations in MATLAB/Simulink. The proposed HTVA method is compared with the standard threshold virtual admittance (TVA) method \cite{rodriguezControlPVGeneration2013} and the voltage information-based virtual admittance (VAv) method which is based on implementing the VIv method in a virtual admittance structure.

\subsection{Test Case Setup}

\begin{figure}
    \centering
    \includegraphics[width=\linewidth]{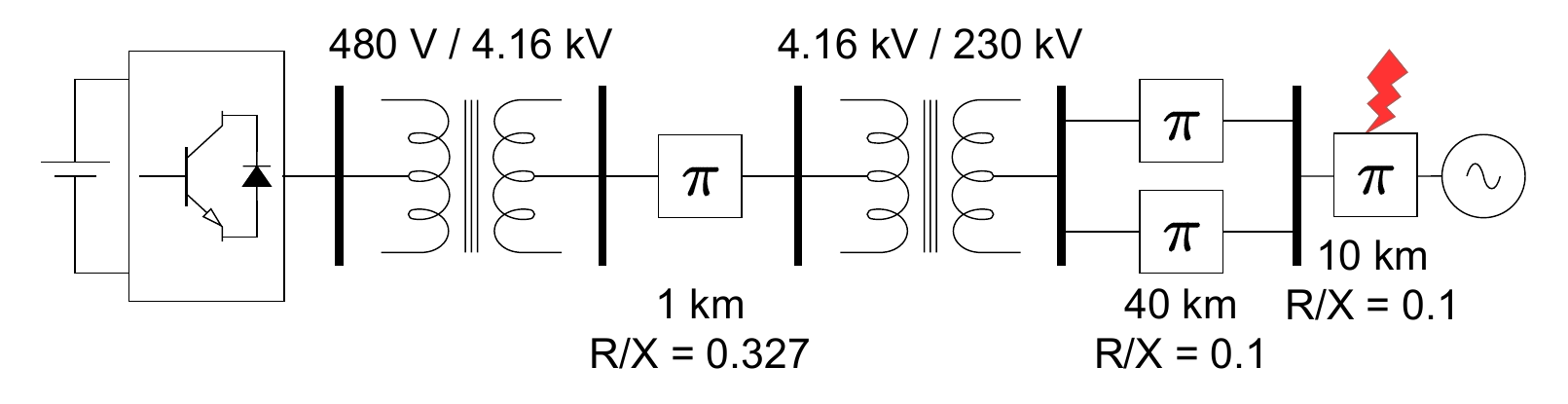}
    \caption{Test system with the GFM inverter connected to an infinite bus through step-up transformers and transmission lines.}
    \label{fig:system} 
\end{figure}

The simulations are carried out in the test system shown in Fig. \ref{fig:system} adopted from \cite{zengHybridThresholdVirtual2023}. The test case contains a droop-controlled GFM inverter switching model linked to an infinite bus via a 1 km medium-voltage line and a 40 km double-circuit high-voltage transmission line, with interfacing step-up transformers. Table I and Table II list the relevant system and control parameters. 

\begin{table}[!t]
\centering
\caption{System and Control Parameters.}
\centering
\begin{tabular}{|l|c|}
\hline
\textbf{Description} & \textbf{Value} \\
\hline
Inverter base power, $S_b$ & 1 MVA \\
Inverter base voltage, $V_b$ & 480 V \\
System base frequency, $f_0$ & 60 Hz \\
Filter resistance, $R_f$ & 0.1 p.u. \\
Filter inductance, $L_f$ & 0.156 p.u. \\
Filter capacitance, $C_f$ & 0.023 p.u. \\
P-Droop coefficient, $m_p$ & 0.05 \\
Q-Droop coefficient, $m_q$ & 0.05 \\
Active power set-point, $P^*$ & 0.5 p.u.\\
Re-active power set-point, $Q^*$ & 0.5 p.u.\\
Cut-off frequency, $\omega_f$ & $2\pi6\ rad/s$ \\
PI coefficients of current controller, $K_I$ and $K_P$ & 2.4 p.u., 1.1 p.u. \\
\hline
\end{tabular}
\end{table}

\begin{table}[!t]
\centering
\caption{Current Limiting Parameters.}
\centering
\begin{tabular}{|l|c|}
\hline
\textbf{Description} & \textbf{Value} \\
\hline
Maximum current, $I_{max}$ & 1.2 p.u. \\
Threshold current, $I_{th}$ & 1 p.u. \\
Steady-state X/R ratio, $\sigma_{SS}$ & 8 \\
Transient X/R ratio, $\sigma_{TR}$ & 0.1 \\

\hline
\end{tabular}
\end{table}

The case studies are outlined as follows:

\begin{itemize}
    \item \textit{Test Case 1. Three-Phase Fault:} At $t=0\ s$, a three-phase ground fault is applied at the distribution line as marked in Fig. \ref{fig:system}. The voltage at the point of common coupling (PCC) drops to $0.4\ p.u.$. The fault is cleared after $0.8\ s$. 
       
    \item \textit{Test Case 2. Grid Voltage Phase Jump:} At $t=0\ s$, a phase shift of $-110\degree$ is applied to the grid voltage.
   
\end{itemize}

\subsection{Test Case 1. Three-Phase Fault}

\begin{figure}[!t]
        \centering
        \includegraphics[width=\linewidth]{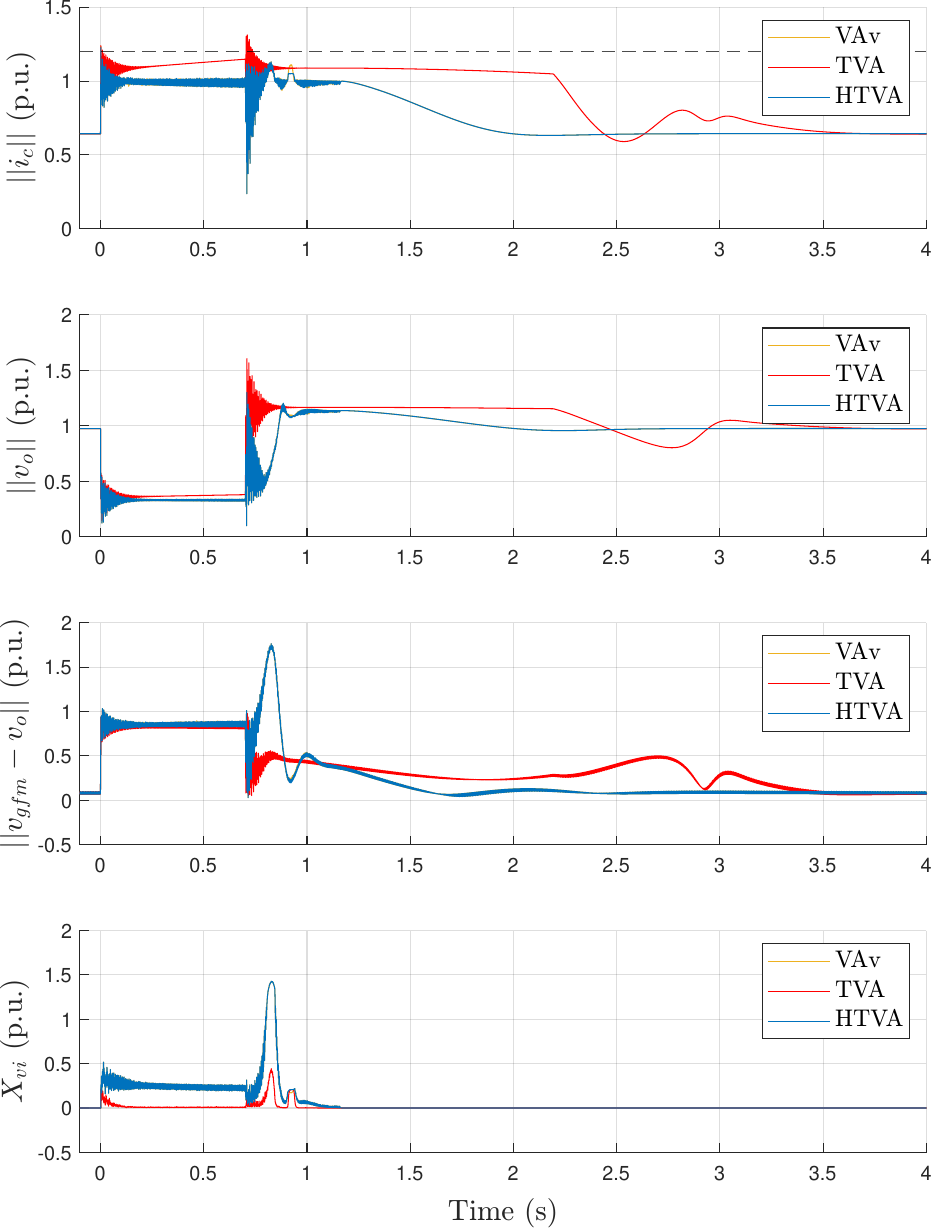}
        \caption{Comparison of transient responses for the VIv, TVI, and proposed HTVI methods, showing $\norm{i_c}$, $\norm{v_o}$, $\norm{v_{gfm} - v_{o}}$, and $X_{vi}$ during the three-phase fault event.}
        \label{fig:fault}
    \end{figure}

Fig. \ref{fig:fault} presents the changes in the magnitudes of $i_c$, $v_o$, $v_{gfm} - v_o$ and $X_{vi}$ with TVA, VAv and XTVA-based current limiting employed. $I_{max} = 1.2\ p.u.$ is shown by a dashed line. It is observed that all three methods can limit the fault current to acceptable levels. As the fault is not at the PCC, $\norm{v_{gfm} - v_{o}} < V_n$, resulting in the current being limited to below $I_{max}$. After the fault is cleared, the systems reach steady-state within a few seconds, while the current is restricted below the allowed limit. 
Notably, the VAv method outperforms the TVA method at fault clearance; the TVA method current goes over $I_{max}$ for an instant, while the VAv method limits the current successfully. This is due to $X_{vi}$ being larger for the VAv method than the TVA method at that instant. Since the HTVA method follows the maximum $X_{vi}$, the hybrid method can limit the current successfully.

\subsection{Test Case 2. Phase Jump}

 \begin{figure}[!t]
        \centering
        \includegraphics[width=\linewidth]{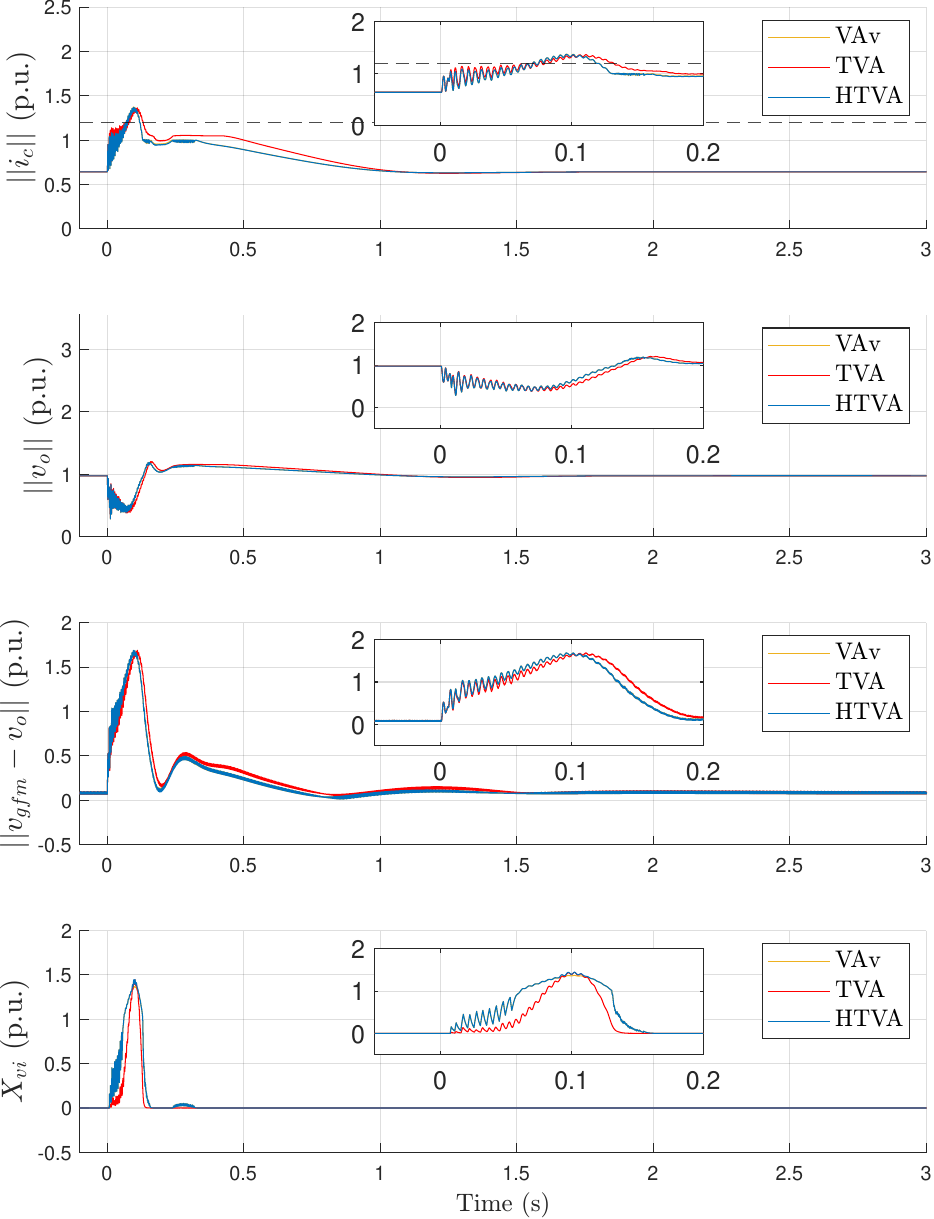}
        \caption{Comparison of transient responses for VIv, TVI, and HTVI methods, showing $\norm{i_c}$, $\norm{v_o}$, $\norm{v_{gfm} - v_{o}}$, and $X_{vi}$ during the phase jump.}
        \label{fig:phase}
    \end{figure}

Fig. \ref{fig:phase} presents the responses of $i_c$, $v_o$, $v_{gfm} - v_o$ and $X_{vi}$ for the phase jump scenario. The phase jump applied at the grid voltage propagates to the filter terminal, causing $\norm{v_{gfm} - v_{o}} > V_n$. 

It is seen that both the VAv and the TVA methods have very similar responses. Both methods have the current go slightly above $I_{max}$ for a small duration. The VAv method brings back the current marginally faster than the TVA method. Accordingly, the proposed HTVA method follows the VAv method and performs slightly better than the standard method. Essentially, due to delays introduced by the virtual admittance, the virtual impedance values do not strictly follow the proposition from \cite{zengHybridThresholdVirtual2023}. 

The case studies demonstrate that the proposed HTVA method can limit the phase current effectively in both three-phase fault and phase jump scenarios. Although the improvements are not as pronounced as in the dual-loop virtual impedance structure, the HTVA method in the single-loop structure still performs better than the standard TVA method.  

\section{Conclusion}

This paper proposes a new HTVA current limiting method for GFM inverters. By combining the strengths of traditional TVI and VIv methods and applying them in a virtual admittance structure, the proposed approach introduces a phase information-aware virtual admittance-based current limiting method. The HTVA method's performance is evaluated through simulations, demonstrating its ability to limit fault currents effectively and transition smoothly to steady-state operation in both three-phase ground faults and phase jump scenarios. This work has potential applications in GFM inverter structures with the single-loop structure. Future work will extend the proposed method to GFM structures that employ alternative structures, such as the single-loop voltage-magnitude controller. 

\IEEEpubidadjcol

\bibliographystyle{IEEEtran}
\bibliography{zaid}
\let\mybibitem\bibitem

\end{document}